\documentclass{osa-article}

\def\arXiv{1}

\articletype{Research Article}

\usepackage[utf8]{inputenc}
\usepackage[english]{babel}

\usepackage{amsmath}
\usepackage{amsfonts}

\usepackage{siunitx}
\newcommand*\chem[1]{\ensuremath{\mathrm{#1}}}      

\usepackage{hyperref}

\usepackage{threeparttable}
\newcolumntype{Y}{>{\centering\arraybackslash}X}
\usepackage{hhline}
\usepackage{enumitem}

\usepackage{xcolor}

\newcommand*\comment[1]{\iffalse#1\fi}
\newcommand\figureScaling{0.9}

\if\arXiv1
    \journal{osajournal}
    \renewenvironment{abstract}
    {\vskip1pc\noindent\textbf{Abstract:\space}}
    {\\[12pt]\noindent{}\par\vskip12pt}
\fi

\begin{document}

\title{Tunable Fiber Fabry-Perot Cavities with High Passive Stability}

\author{Carlos Saavedra,\authormark{1,2,*} Deepak Pandey,\authormark{1,$\dagger$} Wolfgang Alt,\authormark{1} Hannes Pfeifer\authormark{1} and Dieter Meschede\authormark{1}}
\address{\authormark{1}Institut für Angewandte Physik, Universität Bonn, Wegelerstr. 8, 53115 Bonn, Germany\\
\authormark{2}División de Ciencias e Ingenierias Universidad de Guanajuato, México\\}

\email{\authormark{*}carlos.salazar@iap.uni-bonn.de,\authormark{$\dagger$}d.pandey@iap.uni-bonn.de}


\homepage{http://quantum-technologies.iap.uni-bonn.de/} 


\begin{abstract}
We present three high finesse tunable monolithic fiber Fabry-Perot cavities (FFPCs) with high passive mechanical stability. The fiber mirrors are fixed inside slotted glass ferrules, which guarantee an inherent alignment of the resonators. An attached piezoelectric element  enables fast tuning of the FFPC resonance frequency over the entire free-spectral range for two of the designs. Stable locking of the cavity resonance is achieved for feedback bandwidths as low as $\SI{20}{\milli\hertz}$, demonstrating the high passive stability. At the other limit, locking bandwidths up to $\SI{27}{\kilo\hertz}$, close to the first mechanical resonance, can be obtained. 
 The root-mean-square frequency fluctuations are suppressed down to $\sim\SI{2}{\percent}$ of the cavity linewidth. Over a wide frequency range, the frequency noise  is dominated by  the thermal noise limit of the system's mechanical resonances. The demonstrated small footprint devices can be used advantageously in a broad range of applications like cavity-based sensing techniques, optical filters or quantum light-matter interfaces.
\end{abstract}

\section{Introduction}
Miniaturized Fabry-Perot cavities are based on mirrors that are directly fabricated onto the end facets of optical fibers. They have emerged as a versatile optical resonator platform during the last decade \cite{hunger2010fiber, gallego2016high}. They can provide small mode-volumes for high field concentrations in order to enhance light-matter interaction, while at the same time being inherently fiber coupled. Hence, they have quickly evolved into a standard platform to optically interface quantum emitters like atoms \cite{gallego2018strong, colombe2007strong, Macha2020,brekenfeld2020quantum}, ions \cite{steiner2013single,Takahashi2020}, molecules \cite{toninelli2010scanning}, and solid state systems like quantum dots \cite{miguel2013cavity} or NV-centers \cite{albrecht2013coupling}. Furthermore, they have been successfully used in cavity-optomechanical experiments involving a membrane inside the fiber Fabry-Perot cavity (FFPC) \cite{flowers2012fiber}, for sensing of strain \cite{jiang2001simple} and vibration \cite{garcia2010vibration}, for cavity-enhanced microscopy \cite{Mader2015}, and they have been proposed as a fiber-coupled system for optical filtering \cite{ott2016millimeter}. 

 The fiber mirrors that constitute an FFPC are created by laser ablation and a subsequent high-reflective coating of the end-facets of an optical fiber\cite{hunger2010fiber,Uphoff_2015}. Assembling an FFPC typically requires an iterative adjustment of the two opposing fiber mirrors, using three translation and two angular degrees of freedom, to achieve optimal cavity alignment \cite{Brandstatter2013, hunger2010fiber}. In order to control cavity birefringence also rotary adjustment of the fibers is needed\cite{Uphoff_2015,Garcia18}. After alignment, the fibers are glued to their respective holders, e.g. v-grooves. These are mounted on piezo-electric elements, attached to a common base, to enable tuning of the cavity resonance \cite{gallego2016high,Brandstatter2013}.
 
 Such conventional FFPC realizations easily pick up low frequency acoustic noise due to the large distance of the fibers from the common base. Moreover, fiber tips protruding beyond their holders into the free space introduce additional noise due to bending modes. 
 In order to stabilize these cavity systems an electronic locking scheme with feedback bandwidths of the order of several tens of kHz is required \cite{gallego2016high,janitz2017high,brachmann2016photothermal}. An alternative implementation, where the fiber mirrors are fixed inside a common glass ferrule, was demonstrated in \cite{gallego2016high}. This reduces the complexity of the assembling process and at the same time increases the passive stability. The monolithic FFPC in \cite{gallego2016high}, however,  had a small scan range via slow thermal tuning only, and hence was limited in its applicability.

Here we present the fabrication, characterization of optical properties, and locking characteristics of three monolithic FFPCs. The FFPC devices use slotted glass ferrules glued to a piezoelectric element for fast scanning over the entire free-spectral range of the resonators. The fiber mirrors are inherently aligned by the guide provided from the glass ferrule. Due to the high passive stability, feedback bandwidths as low as $\SI{20}{m\hertz}$ are sufficient to lock the cavity resonance to an external laser under laboratory conditions. Fast piezo tuning allows for feedback bandwidths up to $\SI{27}{\kilo\hertz}$ for tight stabilization of the cavity resonance. 

\section{FFPC  design and fabrication}
\begin{figure}[htbp]
    \centering
    \includegraphics[scale=\figureScaling]{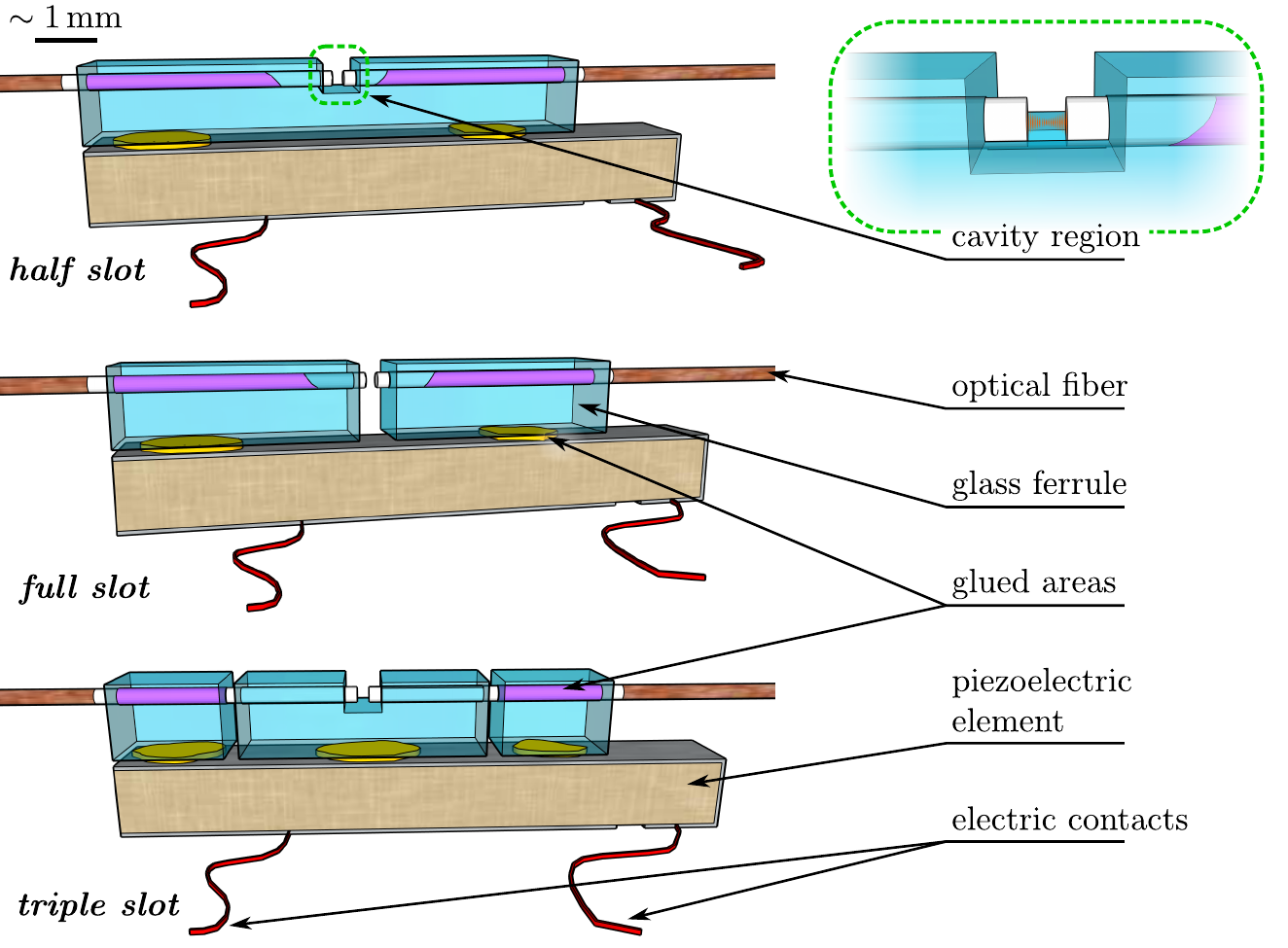}
    \caption{Designs and components of the three FFPCs. The optical cavities are formed by concave dielectric mirrors on the opposing end-facets of the optical fibers (diameters exaggerated by a factor of 2) at the center of the structures. The fibers are glued into the glass ferrules, which are in turn glued to piezoelectric elements for tuning the cavity resonances.
    }
    \label{fig:generalGeometry}
\end{figure}

In Fig.~\ref{fig:generalGeometry} we show the geometry of three FFPC devices based on glass ferrules. These commercially available ferrules ($\SI{8}{\milli \meter}\times \SI{1.25}{\milli \meter} \times\SI{1.25}{\milli \meter}$) are made out of fused silica with  $\SI{131}{\micro\meter}$ nominal inner diameter of the bore. The different slotting patterns of the ferrules are cut using a diamond-plated wire saw. Complete slots are finished only after gluing the ferrule onto the piezo-element to maintain the precise alignment of the bores of the segmented blocks (details see Suppl. Sec.\,1). The electrodes of the piezo-element are connected to copper wires by means of a conductive glue.  
 
In the next step, the optical cavity is formed by inserting the fiber mirrors into the ferrule. The opposing end facets are placed at the center area of the ferrule to form a cavity with length $L_\text{cavity}$ according to the desired free spectral range $\nu_\text{FSR} = c/2\, L_\text{cavity}$ with $c$ the speed of light. In the half and triple slot FFPC the central slot does not fully cut through the bore of the ferrule, such that the fiber tips rest at the base of the bore as shown in the inset of Fig.~\ref{fig:generalGeometry}. In the full slot design, the slotting gap is kept narrow ($\approx\SI{250}{\micro\meter}$) such that there is only a small protrusion of the fiber tips into the free space. This constrains the bending motion of the fiber tips.

To find the optical resonance of the cavity, one of the fiber mirrors is scanned using a piezo-driven translation stage while observing the cavity reflection of an incident probe laser. The  birefringence of the cavity, due to  ellipticity of the mirrors, is reduced by rotating one of the fibers about the cavity axis \cite{Garcia18}. The guide provided by the ferrule significantly simplifies the cavity alignment by requiring only one translation and one rotation degree of freedom, as long as the used fiber mirrors only show small decentrations from the fiber axis. In the last step, the fibers are glued into the ferrule, while keeping the cavity approximately resonant with a laser at the target wavelength. A small amount of UV-curable glue is applied to the fiber where it enters the ferrule. Capillary forces let the glue flow into the space between fiber and ferrule wall, where it is subsequently hardened by UV-light illumination.

The piezo-element is a rectangular ceramic block of dimensions 
$\SI{10}{\milli\meter}\times\SI{1}{\milli\meter}\times\SI{1}{\milli\meter}$ such that it fits the ferrule (Fig.~\ref{fig:generalGeometry}). Applying a voltage to the piezo-element causes a longitudinal displacement of the fiber end facets with respect to each other. The resulting length change $\Delta L_\text{cavity}$, tunes the optical resonance frequencies ($\Delta\nu_\text{scan} = 2\, \nu_\text{FSR} \, \Delta L_\text{cavity} / \lambda$). For fully slotted designs, the tuning range depends on the spacing of the glue points on the piezo element and the length tuning is approximately given by the expansion of the unloaded piezo-element between the glue points. The distance between relevant glue contact points for the full (triple) slot design is $\approx\SI{2.7}{\milli\meter}$ ($\SI{7.4}{\milli\meter}$), translating into a length variation of $\Delta L_\text{cavity}$ $\approx\SI{0.5}{\micro\meter}$ ($\SI{1.5}{\micro\meter}$) for a $\SI{1}{\kilo\volt_{pp}}$ piezo voltage range.
The half slot design shows the smallest tuning range since the length change of the cavity is mediated by the elastic deformation of the stiff ferrule. The measured tuning ranges for the three FFPCs are listed in Tab.~\ref{tab:ffpcOverview}.
\begin{figure}[htbp]
    \centering
     \includegraphics[scale=\figureScaling]{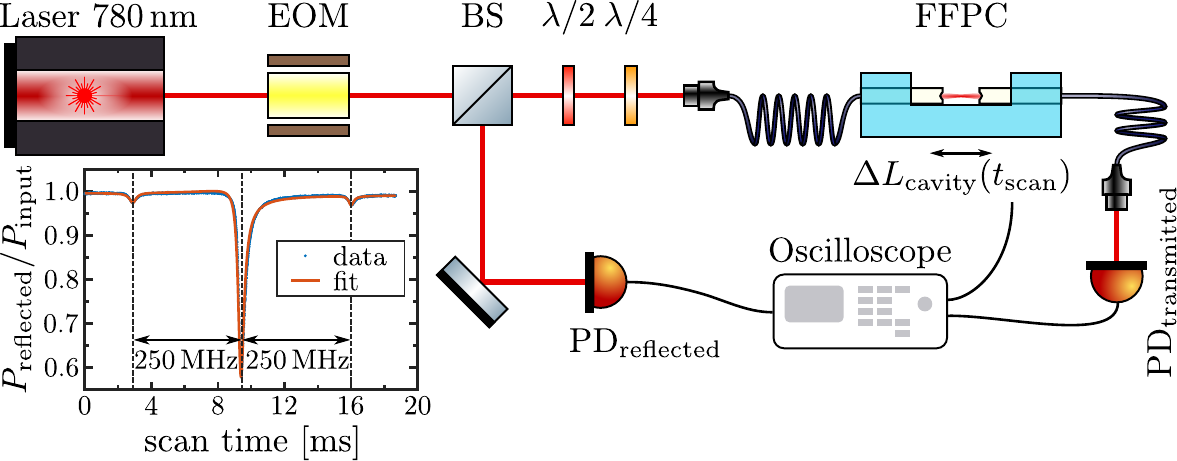}
    \caption{Setup for characterizing FFPCs. The reflected and transmitted laser light power from the FFPC is monitored by the two photodiodes (PD) while the FFPC length $L_\text{cavity}$ is scanned. The waveplates ($\lambda/2$ and $\lambda/4$) before the input fiber are used to investigate the polarization mode splitting of the cavity. The calibration of scan time $t_\text{scan}$ to frequency is achieved by modulating sidebands onto the laser tone using an electro-optic modulator (EOM). BS represents a  non-polarizing beam splitter. (sketch uses \cite{componenentLibraryInkscape})}
    \label{fig:measSetup}
\end{figure}

\section{Optical characterization}
For the characterization of the completed FFPCs, the light from a $\SI{780}{\nano\meter}$ narrow linewidth ($\sim \SI{200}{\kilo \hertz}$) laser  is sent through an electro-optic modulator (EOM) that is driven at $\SI{250}{\mega\hertz}$ to add sidebands to the laser carrier frequency (see Fig.~\ref{fig:measSetup}). A combination of half- and quarter-wave plates is used to control the polarization state of the light before being coupled into the optical fiber that leads to the FFPC. The reflected (transmitted) light from the cavity is directed onto a photodiode PD$_\text{reflected}$ (PD$_\text{transmitted}$). The cavity length is scanned by driving the piezo-element. To measure the cavity linewidth, 
the EOM-generated sidebands are used as frequency markers (see inset in Fig.~\ref{fig:measSetup}).
\begin{figure}[htbp]
    \centering
    \includegraphics[scale=\figureScaling]{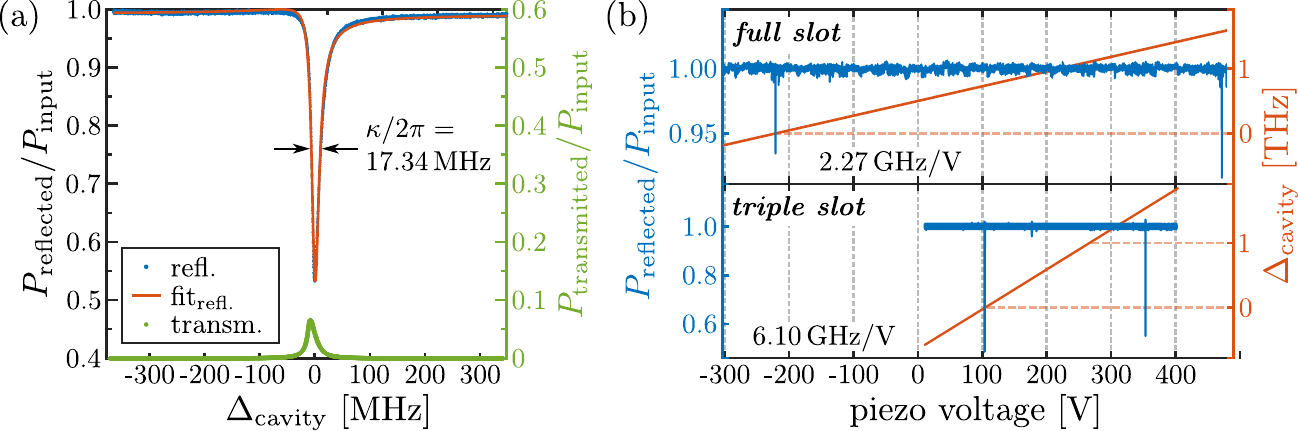}
        \caption{(a) Reflected  and transmitted  power fraction in an exemplary cavity scan of the half slot FFPC with Lorentzian and dispersive fit \cite{gallego2016high}. (b) Scan of the full free spectral range (FSR) for full slot and triple slots FFPCs. 
         Detailed FFPC properties are listed in Tab.~\ref{tab:ffpcOverview}.}
    \label{fig:opticalScans}
\end{figure}

The fiber mirrors used here have a transmission of $T\approx\SI{15}{ppm}$ and absorption of $A\approx\SI{25}{ppm}$ of the dielectric mirror coating. The concave mirror shape on the fiber end facet was created via \chem{CO_2}-laser ablation with a radius of curvature  of $\approx \SI{180}{\micro\meter}$. To experimentally determine the linewith and finesse of the FFPCs, the measured reflection signal is fitted using a sum of a Lorentzian and its corresponding dispersive function \cite{gallego2016high,Bick2016} as shown in Fig.~\ref{fig:opticalScans}~(a). 
In this example of the half slot FFPC, the free spectral range of the $\SI{93}{\micro\meter}$ long cavity is $\nu_\text{FSR} = c/2L_\text{cavity} \sim \SI{1.61}{\tera\hertz}$ yielding a maximum Finesse of $\mathcal{F}\approx \SI{93e3}{}$ from the extracted linewidth of $\kappa/2\pi = (17.34 \pm 0.014)\, \SI{}{\mega\hertz}$. The finesse, the FSR and the linewidths for the three FFPCs are listed in Tab.~\ref{tab:ffpcOverview}. While the half slot FFPC shows a moderate tunability of $\SI{0.13}{\giga\hertz/\volt}$ due to the stiff ferrule geometry, the full and triple slot FFPCs exhibit large tunabilities of $\SI{2.27}{\giga\hertz/\volt}$ and $\SI{6.1}{\giga\hertz/\volt}$ respectively, enabling full FSR scans at moderate voltages as shown in Fig.~\ref{fig:opticalScans} b.

Apart from the piezo-electric tuning, the optical cavity resonance is also sensitive to the ambient temperature \cite{gallego2016high}, which can be used for tuning by $\approx\SI{8}{\giga\hertz/\kelvin}$ (Suppl. Sec.\,5). 
\begin{table}[ht]
\caption{Overview of the optical and locking characteristics of the three FFPCs}
\begin{center}
\begin{threeparttable}
\begin{tabular}{ |p{3.9cm}|p{2.5cm}|p{2.5cm}|p{2.5cm}|}
\hline
Property & Half slot FFPC & Full slot FFPC & Triple slot FFPC \\
\hline
Finesse & 93000 &61000 &99000\\
Linewidth (FWHM) & $\SI{17}{\mega\hertz}$ & $\SI{27}{\mega\hertz}$ & $\SI{16}{\mega\hertz}$\\
FSR & $\SI{1.61}{\tera\hertz}$ & $\SI{1.62}{\tera\hertz}$ & $\SI{1.61}{\tera\hertz}$\\
Scan range$^*$  & $\SI{0.13}{\tera\hertz}$ & $\SI{2.27}{\tera\hertz}$ & $\SI{6.10}{\tera\hertz}$ \\
Max. locking bandwidth & $\SI{20}{\kilo\hertz}$ & $\SI{25}{\kilo\hertz}$ & $\SI{27}{\kilo\hertz}$ \\
Mechanical resonance $^{\#}$  & $\SI{56}{\kilo\hertz}$ & $\SI{34}{\kilo\hertz}$ & $\SI{32}{\kilo\hertz}$ 
\\
Min. locking bandwidth & $\SI{20}{\milli\hertz}$ & $\SI{65}{\milli\hertz}$ & $\SI{110}{\milli\hertz}$ \\
Locked freq. noise (rms)$^{\S}$ & $\SI{0.37}{\mega\hertz}$ & $\SI{0.40}{\mega\hertz}$ & $\SI{0.64}{\mega\hertz}$ \\
\hline
\end{tabular}
\begin{tablenotes}
\item[$^*$] $\SI{1}{\kilo\volt_{pp}}$\,;\,\,\,\, $^{\#}$ Lowest frequency mode\,;\,\,\,\,$^{\S}$ Integrated for $\SI{10}{\hertz}-\SI{1}{\mega\hertz}$ 
\end{tablenotes}
\end{threeparttable}
\end{center}
\label{tab:ffpcOverview}
\end{table}
\begin{figure}[htbp]
    \centering
    \includegraphics[scale=\figureScaling]{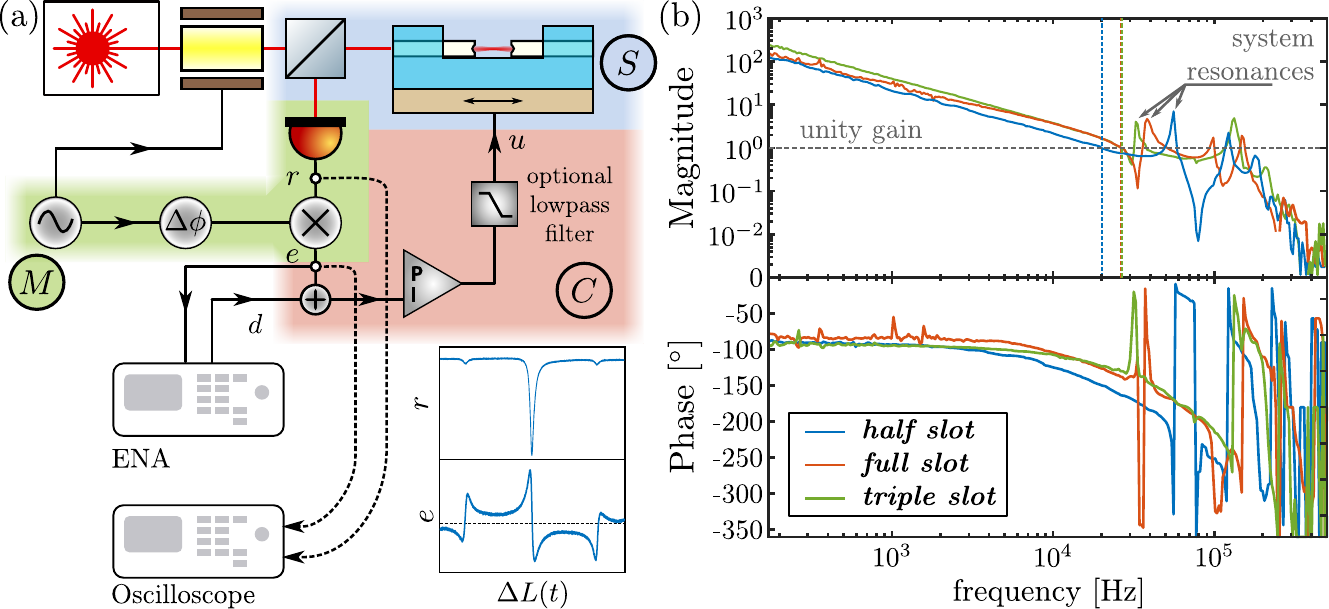}
    \caption{(a) Schematic of the PDH-locking setup for investigating the feedback bandwidth and stability of monolithic FFPCs. The closed feedback system consists of the FFPC device $(S)$, the PDH mixer setup $(M)$, and the feedback controller $(C)$. The input to the PI-controller is the PDH error signal $e$. The output voltage $u$ is applied to the piezo.  The gain of the  PI-controller can be adjusted to explore different locking bandwidths. To measure the frequency response of the closed-loop circuit a frequency sweep signal $d$ from the electrical network analyser (ENA) can be added to $e$. (b) The plots show the magnitude and phase of the full system ($CSM$) transfer function for the maximum achieved bandwidths (dash-dotted vertical lines) of the three designs listed in Tab.~\ref{tab:ffpcOverview}. 
    Sketch uses \cite{componenentLibraryInkscape}.}
    \label{fig:lockingBandwidth}
\end{figure}
\begin{figure}[htbp]
    \centering
    \includegraphics[scale=\figureScaling]{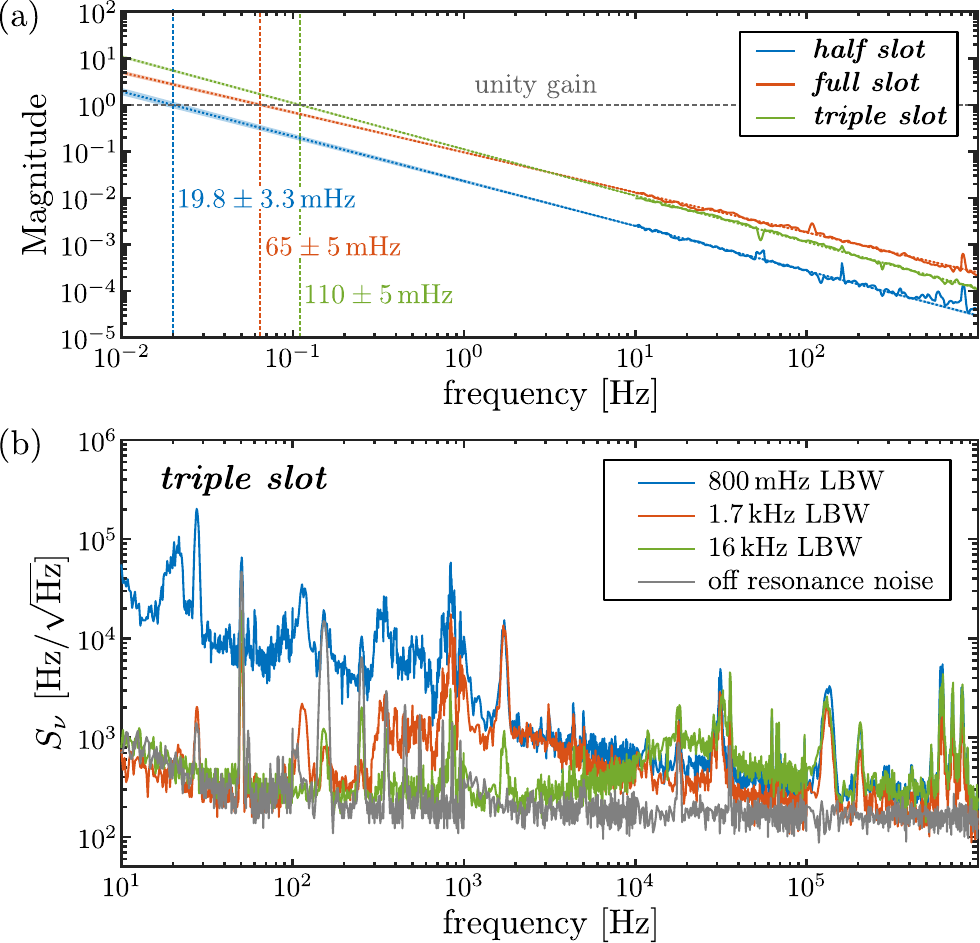}
    \caption{(a) Magnitude of the closed-loop-gain ($CSM(\nu)$) for the three FFPCs with small locking bandwidths (LBW).
    The intersections of the measured transfer functions with the unity gain line are the values corresponding to the low LBW. 
     (b) Measurements of the frequency noise spectral density $S_\nu$ for the triple slot FFPC for three different LBWs. The off resonance noise curve corresponds to the detection noise limit, measured when the cavity is unlocked and far-off resonance.
    }
    \label{fig:lowlocking_bandwidth}
\end{figure}

\section{Cavity locking}
The passive and active stability for the three  FFPCs  is analysed by investigating the closed-loop locking characteristics, where locking refers to an active stabilization of a cavity resonance to a narrow linewidth ($\sim 200$ kHz) laser. We use the standard Pound-Drever-Hall (PDH) locking technique \cite{drever1983laser}, using the EOM-generated sidebands. The schematic of the locking setup is shown in Fig.~\ref{fig:lockingBandwidth} (a). 
The PDH error-signal $e$ is retrieved from the  rf mixer by down-converting the output of the photodiode signal $r$ with the $\SI{250}{\mega\hertz}$ reference signal. The error signal is fed into a PI-controller which drives the piezo attached to the cavity assembly. The PI-controller consists of a variable-gain proportional (P) and an integrator (I) control system. The output of the PI-controller can optionally be low-pass filtered by adding a  resistor in series with the piezo.

We analyse the locking performance of the FFPCs based on techniques used in \cite{Reinhardt:17,janitz2017high}. The frequency-dependent gains (transfer functions) of the components in the servo-loop, as depicted in Fig.~\ref{fig:lockingBandwidth} (a), are:  $C$ - the PI-controller including the optional low pass filter, $S$ - the cavity-assembly, and $M$ - the measurement setup consisting of photodiode and mixer. To retrieve the loop gain $CSM(\nu)$, an electronic network analyzer (ENA) is used to add a frequency-swept external disturbance $d$ to the input of the PI-controller, while the error signal is monitored.

 The loop gain $CSM(\nu)$ is deduced from the measured closed-loop response of the error signal as follows
\begin{equation}
   CSM(\nu) = - \frac{A}{A+1}\,,\,\, A =  \frac{e}{d}.
\end{equation}

We directly obtain $A$ from the network analyzer by monitoring the spectrum of the error signal $e$ (Fig.~\ref{fig:lockingBandwidth} (a)). The closed-loop locking bandwidth (LBW) is given by the unity-gain frequency, i.e the lowest frequency for which $\left|CSM(\nu)\right|=1$. This frequency is the upper limit at which the PI-controller can exert an effective feedback against cavity resonance drifts. By changing the gain settings of the PI-controller, this frequency can be adjusted in order to realize different locking bandwidths. 

\subsection{Maximum locking bandwidth}
To find the maximum achievable LBW of the FFPC devices, the I-gain of the PI-controller is set to the highest value which allows for stable locking. As shown in Fig.~\ref{fig:lockingBandwidth} (b), the maximum LBWs for all three FFPCs are a few tens of kHz, limited by mechanical resonances of the FFPC assembly (details Sect. 4.4). The  rms frequency fluctuation for the maximum LBW is similar to the one given in Tab.~\ref{tab:ffpcOverview}.
 \subsection{Stability of the FFPCs}
For quantitatively characterizing the high passive stability, we extract a value for the minimum locking bandwidth of the FFPCs. For this purpose, the I-gain of the PI-controller is set to the lowest possible value which still allows a stable cavity locking. We found that all three cavities can be locked for many hours under laboratory conditions without any acoustic isolation even for LBW at sub-Hertz level.  Fig.~\ref{fig:lowlocking_bandwidth} (a) shows the plots of $\left|CSM(\nu)\right|$ vs frequency at the lowest gain setting. The LBW is extracted from the intersection of the extrapolated loop gain with the unity gain line. The extracted minimum LBWs for the three FFPCs are given in Tab.~\ref{tab:ffpcOverview}, with the lowest measured value of $\SI{20}{\milli\hertz}$
for the half slot FFPC. As more slots are introduced for achieving larger tunability, the value of the lowest LBW increases successively for full slot and triple slot FFPCs. Therefore, the larger tunability of the FFPC is obtained at the cost of a slight reduction of the stability. This is further characterised by the frequency noise spectrum for a locked FFPC as described below.

\subsection{Noise-spectral-density analysis}
To compare the noise characteristics of the FFPCs  under different lock conditions, we measure their frequency noise spectral densities ($S_\nu$), which are derived from the measured noise spectral densities of the error signals. Fig.~\ref{fig:lowlocking_bandwidth} (b) shows $S_\nu$ for the triple slot FFPC at three different locking bandwidths (see Suppl. Sec.\,3 for the other two FFPCs). The integrated rms frequency noise from these measurements amounts to $\nu_\text{rms}=(1.19,0.64,1.28,0.24)\, \SI{}{\mega\hertz}$ for the $\SI{800}{\milli\hertz}$, $\SI{1.7}{\kilo\hertz}$, $\SI{16}{\kilo\hertz}$ locking bandwidths and the off resonance noise respectively. The off resonance noise is measured with an unlocked cavity far away from resonance. Although the locking bandwidth is changed by almost four orders of magnitude,  the value of the integrated rms frequency noise remains similar, demonstrating the inherent high passive stability of the FFPCs.  The slightly higher noise for sub-Hertz locking bandwidth is due to the uncompensated environmental acoustic noise below $\SI{1}{\kilo\hertz}$, while at a LBW of $\SI{1.7}{\kilo\hertz}$ acoustic noise is suppressed. We have also observed that when the locking bandwidth approaches the mechanical resonances, the rms frequency noise increases again due to the excitation of these resonances. 

\subsection{Mechanical resonances}
The most prominent contribution to the measured frequency noise  at higher frequencies is caused by the mechanical resonances of the assemblies starting around $\SI{30}{\kilo\hertz}$, see Fig.~\ref{fig:lowlocking_bandwidth} and Tab.~\ref{tab:ffpcOverview}. Without active damping of these resonances, the minimum achievable noise corresponds to the thermal excitation of these modes at ambient temperature. 

\begin{figure}[htbp]
    \centering
    \includegraphics[scale=\figureScaling]{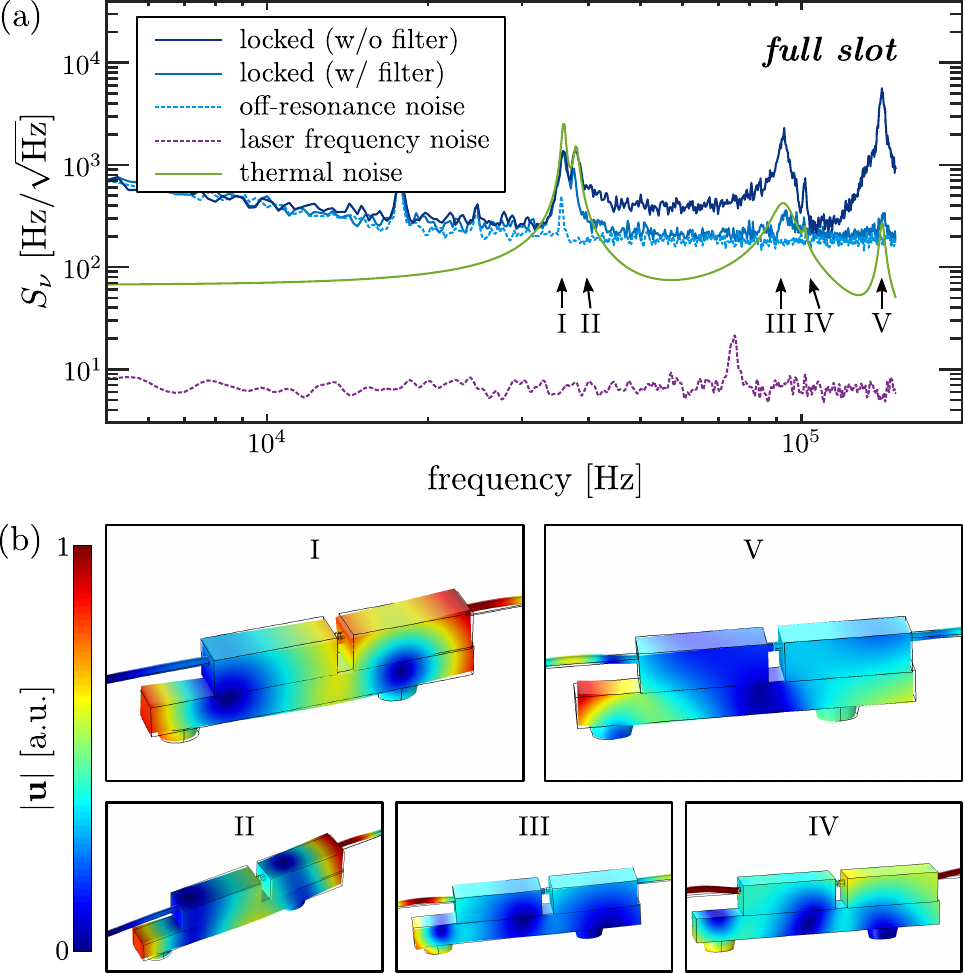}
    \caption{Analysis of the measured FFPC frequency noise induced by mechanical resonances of the system for the full slot FFPC. (a) The measured $S_\nu$ are compared with the laser frequency noise and with the expected noise from thermally excited mechanical resonances of the system. The $S_\nu$ approaches the expected thermal noise limit near the mechanical resonances. Without a low-pass filter, excess electrical noise is coupled to the cavity through the piezoelectric element for some of the modes. (b) Displacement fields of the mechanical resonances included in the model. The two most prominent resonances, I and V, correspond to a bending mode (no piezoelectric coupling) and a longitudinal stretching mode (piezoelectrically coupled) of the structure.}
    \label{fig:mechanicalmodes}
\end{figure}

In order to quantify the thermal noise limit, we performed finite element simulations \cite{COMSOL} of the assembly geometries (see Suppl. Sec.\,4). These are used to extract the displacement fields of the mechanical eigenmodes, their effective masses and optomechanical coupling strengths. Damping was not included in the simulations. The resonances in the recorded noise spectral densities are then attributed to eigenmodes found in the simulation appearing at close-by frequencies. The thermal excitation of the mechanical resonances is calculated from the fluctuation-dissipation theorem using the fitted frequencies and linewidths that parametrize the damping of the modes. Together with the simulated parameters the expected signal in the frequency noise spectrum  \cite{gorodetksy2010determination} can be compared to the measured frequency noise spectral density. An example for this is shown in Fig.~\ref{fig:mechanicalmodes}~(a) for the full slot FFPC. Here, we find five mechanical modes (I-V in Fig.~\ref{fig:mechanicalmodes}~(b)) with substantial coupling to the optical resonance for frequencies up to $\SI{150}{\kilo\hertz}$. The two most prominent modes in the experiment (blue traces), I and V, correspond to a bending and longitudinal stretching motion of the piezo. The calculated thermally induced frequency noise is shown in green in Fig.~\ref{fig:mechanicalmodes}~(a). Since the stretching of the piezo is coupled to the driving voltage by the piezoelectric effect, excess electrical noise from the controller is resonantly coupled to the system at the stretching mode frequency (dark blue curve in Fig.~\ref{fig:mechanicalmodes}~(a)). The excess electrical noise can however be removed by inserting a suitable low pass filter before the piezo (medium blue curve). After inserting the filter, the measured noise is compatible with the theoretically achievable thermal noise limit and the off-resonance detection noise (dashed light blue curve). The frequency noise of the laser does not play a role since it is found to be more than an order of magnitude lower than the detection noise (see Suppl. Sec.\,2).  

\section{Conclusion}
In this paper we have demonstrated a versatile, robust and simple approach for building stable fiber Fabry-Perot cavities with wide frequency tunability and simultaneous high passive stability. The demonstrated high locking bandwidth is feasible due to the lowest mechanical resonance frequencies of the compact devices at few tens of kHz.
Achieving stable locking at tens of mHz feedback bandwidth, which implies that the cavity resonance is stable on a minute time scale for a free running cavity, and reaching the thermal noise limit at higher frequencies, proves the high passive stability of these devices against slow and fast environmental disturbances.
 These compact and inherently fiber coupled cavities can be readily implemented in various applications like cavity-based spectroscopy of gases,  tunable optical filters, and cavity quantum electrodynamics experiments,  which all benefit from highly stable optical resonators.   

By incorporating mode-matched \cite{gulati2017fiber} and millimeters long FFPCs \cite{ott2016millimeter} in our compact devices, stable resonators with linewidths below $\SI{1}{\mega\hertz}$ can be achieved. This will enable the realization of compact gas sensors for airborne applications as well as miniaturized cavity enhanced vapor based light-matter interfaces. Considering ferrules made of ultra-low expansion glass and with cryogenically cooled FFPCs, one can envision miniaturized and portable optical oscillators with extremely high short term stability.

\section*{Acknowledgements}
The authors thank Stephan Kucera for valuable discussions. This work was supported by the Bundesministerium f\"ur Bildung und Forschung (BMBF), projects Q.Link.X and FaResQ. C.S. is supported by a national scholarship from CONACYT, México. W.A. acknowledges funding by the Deutsche Forschungsgemeinschaft (DFG, German Research Foundation) under Germany's Excellence Strategy – Cluster of Excellence Matter and Light for Quantum Computing (ML4Q) EXC 2004/1 – 390534769. 


\begin{thebibliography}{10}
\newcommand{\enquote}[1]{``#1''}

\bibitem{hunger2010fiber}
D.~Hunger, T.~Steinmetz, Y.~Colombe, C.~Deutsch, T.~W. H{\"a}nsch, and
  J.~Reichel, \enquote{A fiber fabry--perot cavity with high finesse,}
  {\protect\JournalTitle{New Journal of Physics}} \textbf{12}, 065038 (2010).

\bibitem{gallego2016high}
J.~Gallego, S.~Ghosh, S.~K. Alavi, W.~Alt, M.~Martinez-Dorantes, D.~Meschede,
  and L.~Ratschbacher, \enquote{High-finesse fiber fabry--perot cavities:
  stabilization and mode matching analysis,} {\protect\JournalTitle{Applied
  Physics B}} \textbf{122}, 47 (2016).

\bibitem{gallego2018strong}
J.~Gallego, W.~Alt, T.~Macha, M.~Martinez-Dorantes, D.~Pandey, and D.~Meschede,
  \enquote{Strong purcell effect on a neutral atom trapped in an open fiber
  cavity,} {\protect\JournalTitle{Physical Review Letters}} \textbf{121},
  173603 (2018).

\bibitem{colombe2007strong}
Y.~Colombe, T.~Steinmetz, G.~Dubois, F.~Linke, D.~Hunger, and J.~Reichel,
  \enquote{Strong atom--field coupling for bose--einstein condensates in an
  optical cavity on a chip,} {\protect\JournalTitle{Nature}} \textbf{450},
  272--276 (2007).

\bibitem{Macha2020}
T.~Macha, E.~Uru{\~n}uela, W.~Alt, M.~Ammenwerth, D.~Pandey, H.~Pfeifer, and
  D.~Meschede, \enquote{Nonadiabatic storage of short light pulses in an
  atom-cavity system,} {\protect\JournalTitle{Physical Review A}} \textbf{101},
  053406 (2020).

\bibitem{brekenfeld2020quantum}
M.~Brekenfeld, D.~Niemietz, J.~D. Christesen, and G.~Rempe, \enquote{A quantum
  network node with crossed optical fibre cavities,}
  {\protect\JournalTitle{Nature Physics}} \textbf{16}, 647–651 (2020).

\bibitem{steiner2013single}
M.~Steiner, H.~M. Meyer, C.~Deutsch, J.~Reichel, and M.~K{\"o}hl,
  \enquote{Single ion coupled to an optical fiber cavity,}
  {\protect\JournalTitle{Physical Review Letters}} \textbf{110}, 043003 (2013).

\bibitem{Takahashi2020}
H.~Takahashi, E.~Kassa, C.~Christoforou, and M.~Keller, \enquote{Strong
  coupling of a single ion to an optical cavity,}
  {\protect\JournalTitle{Physical Review Letters}} \textbf{124}, 013602 (2020).

\bibitem{toninelli2010scanning}
C.~Toninelli, Y.~Delley, T.~St{\"o}ferle, A.~Renn, S.~G{\"o}tzinger, and
  V.~Sandoghdar, \enquote{A scanning microcavity for in situ control of
  single-molecule emission,} {\protect\JournalTitle{Applied Physics Letters}}
  \textbf{97}, 021107 (2010).

\bibitem{miguel2013cavity}
J.~Miguel-S{\'a}nchez, A.~Reinhard, E.~Togan, T.~Volz, A.~Imamoglu, B.~Besga,
  J.~Reichel, and J.~Est{\`e}ve, \enquote{Cavity quantum electrodynamics with
  charge-controlled quantum dots coupled to a fiber fabry--perot cavity,}
  {\protect\JournalTitle{New Journal of Physics}} \textbf{15}, 045002 (2013).

\bibitem{albrecht2013coupling}
R.~Albrecht, A.~Bommer, C.~Deutsch, J.~Reichel, and C.~Becher,
  \enquote{Coupling of a single nitrogen-vacancy center in diamond to a
  fiber-based microcavity,} {\protect\JournalTitle{Physical Review Letters}}
  \textbf{110}, 243602 (2013).

\bibitem{flowers2012fiber}
N.~Flowers-Jacobs, S.~Hoch, J.~Sankey, A.~Kashkanova, A.~Jayich, C.~Deutsch,
  J.~Reichel, and J.~Harris, \enquote{Fiber-cavity-based optomechanical
  device,} {\protect\JournalTitle{Applied Physics Letters}} \textbf{101},
  221109 (2012).

\bibitem{jiang2001simple}
M.~Jiang and E.~Gerhard, \enquote{A simple strain sensor using a thin film as a
  low-finesse fiber-optic fabry--perot interferometer,}
  {\protect\JournalTitle{Sensors and Actuators A: Physical}} \textbf{88},
  41--46 (2001).

\bibitem{garcia2010vibration}
Y.~R. Garc{\'\i}a, J.~M. Corres, and J.~Goicoechea, \enquote{Vibration
  detection using optical fiber sensors,} {\protect\JournalTitle{Journal of
  Sensors}} \textbf{2010} (2010).

\bibitem{Mader2015}
M.~Mader, J.~Reichel, T.~W. H{\"a}nsch, and D.~Hunger, \enquote{A scanning
  cavity microscope,} {\protect\JournalTitle{Nature communications}}
  \textbf{6}, 1--7 (2015).

\bibitem{ott2016millimeter}
K.~Ott, S.~Garcia, R.~Kohlhaas, K.~Sch{\"u}ppert, P.~Rosenbusch, R.~Long, and
  J.~Reichel, \enquote{Millimeter-long fiber fabry-perot cavities,}
  {\protect\JournalTitle{Optics express}} \textbf{24}, 9839--9853 (2016).

\bibitem{Uphoff_2015}
M.~Uphoff, M.~Brekenfeld, G.~Rempe, and S.~Ritter, \enquote{Frequency splitting
  of polarization eigenmodes in microscopic fabry--perot cavities,}
  {\protect\JournalTitle{New Journal of Physics}} \textbf{17}, 013053 (2015).

\bibitem{Brandstatter2013}
B.~Brandst{\"a}tter, A.~McClung, K.~Sch{\"u}ppert, B.~Casabone, K.~Friebe,
  A.~Stute, P.~O. Schmidt, C.~Deutsch, J.~Reichel, R.~Blatt \emph{et~al.},
  \enquote{Integrated fiber-mirror ion trap for strong ion-cavity coupling,}
  {\protect\JournalTitle{Review of Scientific Instruments}} \textbf{84}, 123104
  (2013).

\bibitem{Garcia18}
S.~Garcia, F.~Ferri, K.~Ott, J.~Reichel, and R.~Long, \enquote{Dual-wavelength
  fiber fabry-perot cavities with engineered birefringence,}
  {\protect\JournalTitle{Optics express}} \textbf{26}, 22249--22263 (2018).

\bibitem{janitz2017high}
E.~Janitz, M.~Ruf, Y.~Fontana, J.~Sankey, and L.~Childress, \enquote{High
  mechanical bandwidth fiber-coupled fabry-perot cavity,}
  {\protect\JournalTitle{Optics Express}} \textbf{25}, 20932--20943 (2017).

\bibitem{brachmann2016photothermal}
J.~F. Brachmann, H.~Kaupp, T.~W. H{\"a}nsch, and D.~Hunger,
  \enquote{Photothermal effects in ultra-precisely stabilized tunable
  microcavities,} {\protect\JournalTitle{Optics express}} \textbf{24},
  21205--21215 (2016).

\bibitem{componenentLibraryInkscape}
A.~Franzen, \enquote{{ComponentLibrary},} Licensed under CC BY-NC 3.0,\\
  \url{http://www.gwoptics.org/ComponentLibrary/}.

\bibitem{Bick2016}
A.~Bick, C.~Staarmann, P.~Christoph, O.~Hellmig, J.~Heinze, K.~Sengstock, and
  C.~Becker, \enquote{The role of mode match in fiber cavities,}
  {\protect\JournalTitle{Review of Scientific Instruments}} \textbf{87}, 013102
  (2016).

\bibitem{drever1983laser}
R.~Drever, J.~L. Hall, F.~Kowalski, J.~Hough, G.~Ford, A.~Munley, and H.~Ward,
  \enquote{Laser phase and frequency stabilization using an optical resonator,}
  {\protect\JournalTitle{Applied Physics B}} \textbf{31}, 97--105 (1983).

\bibitem{Reinhardt:17}
C.~Reinhardt, T.~M{\"u}ller, and J.~C. Sankey, \enquote{Simple delay-limited
  sideband locking with heterodyne readout,} {\protect\JournalTitle{Optics
  Express}} \textbf{25}, 1582--1597 (2017).

\bibitem{COMSOL}
{COMSOL AB}, \enquote{{COMSOL} {M}ultiphysics\textsuperscript{\textregistered}
  v. 5.2,} www.comsol.com. Stockholm, Sweden.

\bibitem{gorodetksy2010determination}
M.~Gorodetksy, A.~Schliesser, G.~Anetsberger, S.~Deleglise, and T.~J.
  Kippenberg, \enquote{Determination of the vacuum optomechanical coupling rate
  using frequency noise calibration,} {\protect\JournalTitle{Optics express}}
  \textbf{18}, 23236--23246 (2010).

\bibitem{gulati2017fiber}
G.~K. Gulati, H.~Takahashi, N.~Podoliak, P.~Horak, and M.~Keller,
  \enquote{Fiber cavities with integrated mode matching optics,}
  {\protect\JournalTitle{Scientific Reports}} \textbf{7}, 1--6 (2017).

\end{thebibliography}

\end{document}